# Ontology-based annotation and analysis of COVID-19 phenotypes

Yang Wang [a,b], Fengwei Zhang [a], Hong Yu [a,b], Xianwei Ye [a,b], Yongqun He [c,1]

[a] *Guizhou University Medical College, Guiyang, Guizhou 550025, China.*

[b] *Department of Pulmonary and Critical Care Medicine, Guizhou Provincial People's Hospital and NHC Key Laboratory of Immunological Diseases, People's Hospital of Guizhou University, Guiyang, Guizhou 550002, China.*

[c] *University of Michigan Medical School, Ann Arbor, MI 48109, USA.*

**Abstract.** The epidemic of COVID-19 has caused an unpredictable and devastated disaster to the public health in different territories around the world. Common phenotypes include fever, cough, shortness of breath, and chills. With more cases investigated, other clinical phenotypes are gradually recognized, for example, loss of smell, and loss of tastes. Compared with discharged or cured patients, severe or died patients often have one or more comorbidities, such as hypertension, diabetes, and cardiovascular disease. In this study, we systematically collected and analyzed COVID-19-related clinical phenotypes from 70 articles. The commonly occurring 17 phenotypes were classified into different groups based on the Human Phenotype Ontology (HPO). Based on the HP classification, we systematically analyze three nervous phenotypes (loss of smell, loss of taste, and headache) and four abdominal phenotypes (nausea, vomiting, abdominal pain, and diarrhea) identified in patients, and found that patients from Europe and USA turned to have higher nervous phenotypes and abdominal phenotypes than patients from Asia. A total of 23 comorbidities were found to commonly exist among COVID-19 patients. Patients with these comorbidities such as diabetes and kidney failure had worse outcomes compared with those without these comorbidities.

## 1. Introduction

The COVID-19 pandemic had resulted in a major loss worldwide. In the US alone, the pandemic had caused over 1.6 million confirmed cases and near 10 thousand deaths by May 25, 2020. The cause of the disease is SARS-CoV-2, a single-positive RNA virus. It belongs to the order *Nidovirales*, family *Coronaviridae*, and subfamily

---

[1] Yongqun Oliver He, Corresponding author, University of Michigan Medical School, Ann Arbor, MI, USA. E-mail: yongqunh@med.umich.edu.

*Coronavirinae* [4]. According to the literature and newspapers reporting, the clinical symptoms in different areas are overall stable but also show differences, which we often ignored. The common symptoms include fever, cough, shortness of breath or difficulty breathing, and chills [12]. However, physicians and pathologists have recognized other clinical phenotypes, for example, loss of smell, and loss of tastes. In many cases, the symptoms of the digestive system become the primary phenotypes [12]. The early recognition of these symptoms facilitates early diagnosis and treatment, which would bring immeasurable good outcomes to them.

Compared to discharged or cured patients of COVID-19, severe or died patients are more likely to have comorbidities. Comorbidity here refers to the simultaneous presence of some disease(s) or condition(s) in a patient when COVID-19 occurs. Hypertension, diabetes, cardiovascular diseases are a few examples of the most common comorbidities associated with COVID-19 [1]. However, a systematic investigation and classification of the relations between the comorbidities and disease outcomes have not been reported.

Ontology has played a significant role in standard data representation, classification, integration, and analysis. Many ontologies have been widely used in the field of medicine and pharmacy in recent years. The Human Phenotype Ontology (HPO) is a standardized vocabulary for phenotypic abnormalities in human disease. Each term in the HPO describes a phenotypic abnormality, such as Pneumonia. HPO currently contains over 13,000 terms [6]. It bridges a computational link between genome biology and clinical medicine.

In this study, we systematically annotated COVID-19 clinical phenotypes and comorbidities from journal and preprint articles, and applied HP to classify these phenotypes and examine the internal patterns. Our study also identified many shared and differential phenotype patterns in COVID-19 patients from countries in different regions in the world.

## 2. Methods

### 2.1. Data collection

Peer-reviewed journal articles and preprint bioRxiv and medRxiv articles were searched to identify relevant articles.

### 2.2. Ontology classification

The identified symptoms and comorbidities were mapped to the terms in the Human Phenotype ontology (HPO). Ontobee [13] was used to search the HP terms. Ontofox [16] was used to extract the specific sets of the phenotypes. The Protégé-OWL editor was used for display, classification, and analysis.

### 2.3. Ontology modeling

The mapped and extracted HP terms and subsets were imported to the Coronavirus Infectious Disease Ontology (CIDO) (http://github.com/CIDO-ontology/cido).

## 3. Results

*3.1. Data collection*

We reviewed 70 papers from December 2019 to date in Pubmed and preprint bioRxiv and medRxiv. The patients reported in these papers include countries in Asia (e.g., China, South Korea, Singapore, Japan), Middle East(e.g., Iran), Europe (e.g., Italy, France, Spain, German, Belgium, Switzerland), and North America (the USA). Ten representative articles are provided in Table 1.

**Table 1.** Ten representative articles annotated in this study

| Report time | Country | Cases | Reference (PMID or DOI) |
|---|---|---|---|
| Feb 19, 2020 | China | 140 | PMID: 32077115 |
| Feb 28, 2020 | China | 1099 | PMID: 32109013 |
| Mar 27, 2020 | Iran | 10069 | Doi:https://doi.org/10.1101/2020.03.23.20041889 |
| Apr 17, 2020 | France | 114 | PMID: 32305563 |
| Apr 22, 2020 | America | 5700 | PMID: 32320003 |
| Apr 22, 2020 | Italy | 374 | PMID: 32320008 |
| Apr 24, 2020 | America | 169 | PMID: 32329222 |
| Apr 24, 2020 | America | 1299 | PMID: 32329797 |
| Apr 28, 2020 | Britain | 16749 | Doi:https://doi.org/10.1101/2020.04.23.20076042 |
| Apr 30, 2020 | France、Italy、Spain、Belgium、Switzerland | 1420 | PMID: 32352202 |

*3.2. Ontology-based classification of common phenotypes*

Based on our literature search, we found a large number of COVID-19 case reports and analyses from December 2019 to date in different regions all around the world. A total of 17 common clinical symptoms were found, including fever, cough, shortness of breath or difficulty breathing, chills, repeated shaking with chills, muscle pain, headache, sore throat, new loss of smell or taste etc. HP was used to classify these 17 common symptoms (Figure 1). Overall, these symptoms are located in the abdominal system, nervous system, head, respiratory system, constitutional system, and blood. The nervous system abnormality includes parageusia (loss of taste, HP:0031249), anosmia (loss of smell, HP:0000458), and headache (HP:0002315). The abnormality of the head includes abnormality of face and pharynx. Parageusia and anosmia are abnormality of face phenotypes. Pharingitis belongs to the abnormality of nasopharynx and pharynx.

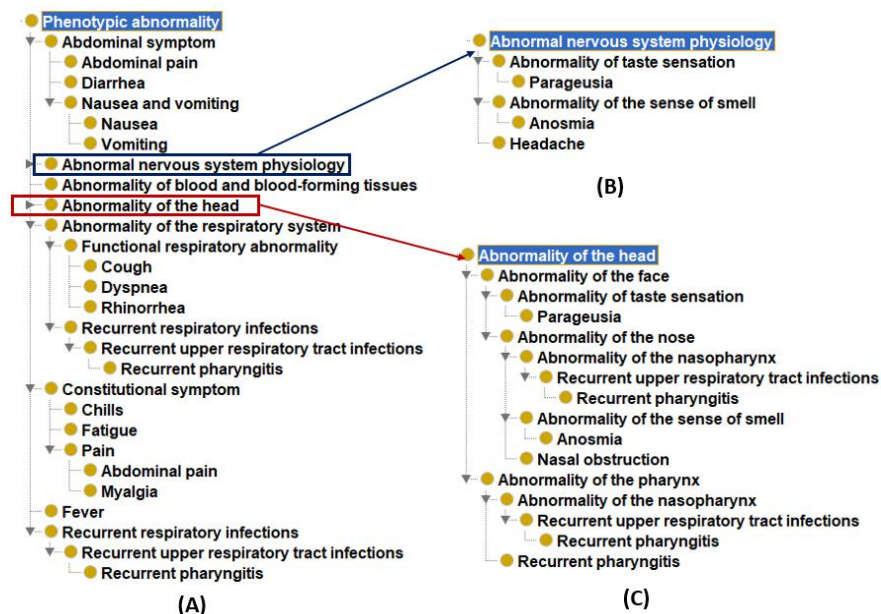
**Figure 1. Common phenotypes of COVID-19 based on HP classification.**

*3.3. Differential profiles of nervous phenotypes in patients from different regions*

Instead of focusing on individual phenotypes, we hypothesized that the analysis of phenotypes as groups would identify new scientific insights. First, we analyzed the group of COVID-19-related nervous system phenotypes, which includes loss of smell (anosmia), loss of taste (parageusia), and headache (Figure 1). The combined analysis of all these three phenotypes as a whole provided us a unique angle to study how the disease affects the nervous system.

As shown in Figure 2, all the three nervous symptom phenotypes in Asian patients appeared low. Ten groups of Chinese patients were analyzed. Among the three phenotypes, headache was relatively common in Chinese patients. However, except one group reporting low level of hyposmia and hypogeusia [11], the other 9 groups did not report any cases of loss of smell and loss of taste. South Korean and Japanese also show the consistent pattern of lower incidence of loss of smell and loss of taste in Figure2.

In contrast, in Iran, Italy, France, German, Spain, USA, there were higher proportions of cases with the loss of smell or tastes. Especially in Europe, there were two large investigation questionnaires [9] about loss of smell and taste in confirmed patients, including many doctors and nurses infected in hospital. According to the multicenter study, the loss of smell was a key symptom in mild-to-moderate COVID-19 patients, the loss of smell was also not associated with nasal obstruction and rhinorrhea. Females and young patients were more susceptible to having the symptoms of smell and taste loss, whereas elderly individuals often presented fever, fatigue and loss of appetite. An obvious correlation between smell and taste disorders was also identified [9].

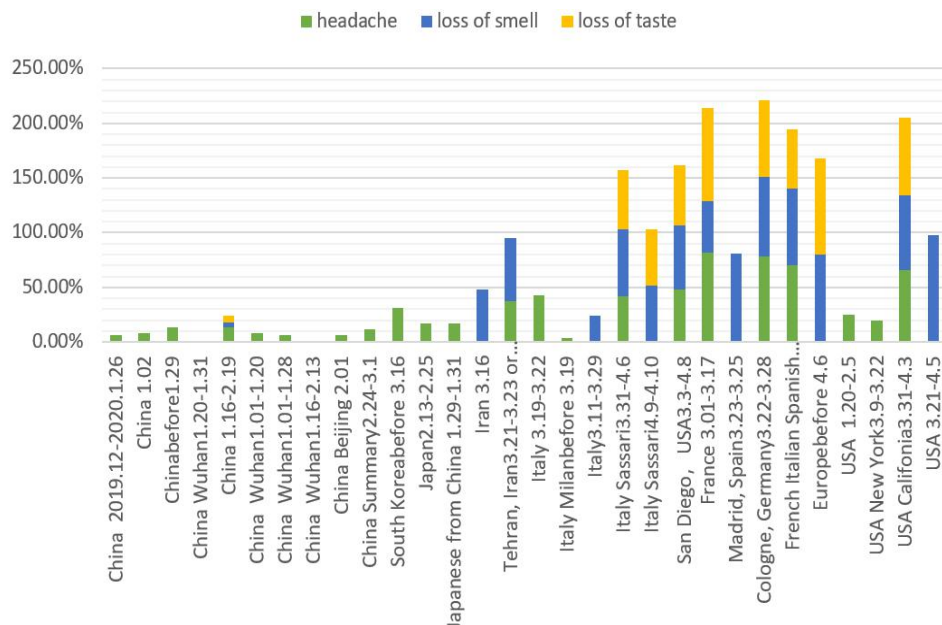

**Figure 2. Cases with three nervous system phenotypes in different countries.** Three symptoms (i.e., headache, loss of smell, and loss of taste) were analyzed. Each symptom has 0-100% of occurrence.

*3.4. Differential profiles of abdominal phenotypes in patients from different regions*

Many mild-moderate COVID-19 patients had gastrointestinal disorders as primary symptoms. The primary abdominal phenotypes include nausea, vomiting, abdominal pain, and diarrhea. We analyzed these four abdominal phenotypes together and compared the cases from different countries and regions.

A retrospective case-control study in New York found that those patients with gastrointestinal symptoms (defined as diarrhea or nausea/vomiting) were significantly more likely to be tested COVID-19 positive than negative (61% vs. 39%, p=0.04) [12]. In 393 patients with COVID-19 in two hospitals in New York, the disease manifestations were in general similar to those in a large case series from China [7]; however, gastrointestinal symptoms appeared to be more common in New York than in China (where these symptoms occurred in 4 to 5% of patients) [5]. We also found that digestive symptoms, especially diarrhea, occurred in almost all countries (Figure 3). However, the accumulated percentages of digestive symptoms were significantly higher in UK, France, New York, California. The time also appears to be a factor. More digestive symptoms were showed in middle-late of March and early April.

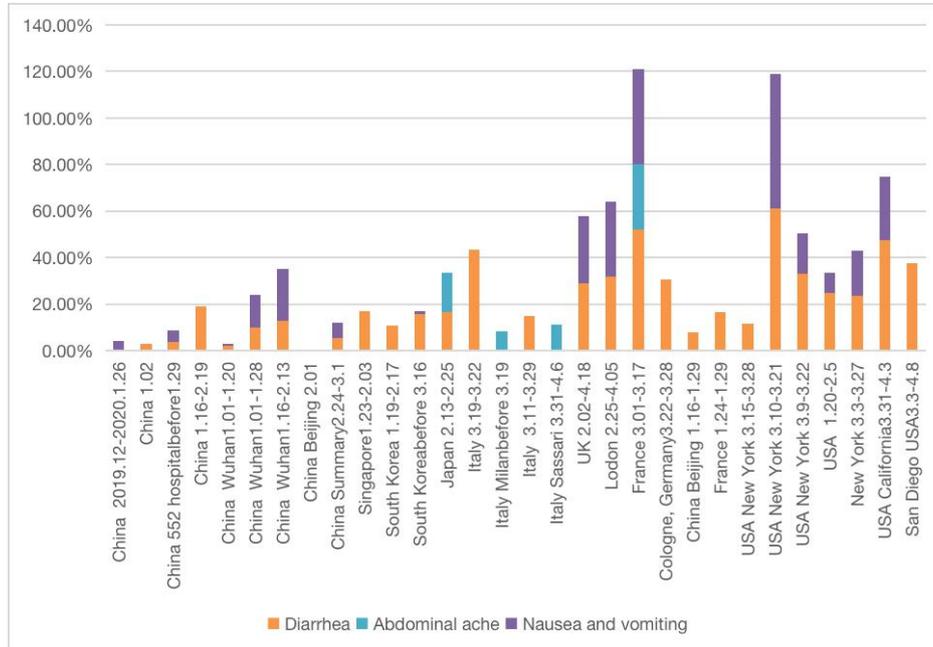

**Figure 3. Cases with four abdominal system phenotypes in different countries.** Four symptoms (i.e., nausea, vomiting, abdominal pain, and diarrhea) were analyzed. Each symptom has 0-100% of occurrence.

*3.5. Ontology-based analysis of the relation between comorbidity vs disease outcomes*

Many factors, such as age, gender, smoking status, have been found to significantly affect disease outcomes. For example, among reported COVID-19 cases, the older men have a higher mortality rate [2]. Comorbidity (i.e., existing medical conditions when patient is infected) is another important risk factor for outcomes. Hypertension, diabetes, cardiovascular diseases, chronic pulmonary diseases are the most common comorbidities. Other complication comorbidities include chronic kidney, hepatitis B/C infection, chronic hepatic failure, cirrhosis, chronic neurological disease (e.g., seizures and dementia), and haematological system disease (e.g., abnormality of blood and blood-forming tissues) [5].

We identified 23 common comorbidities and used HP to classify these comorbidities (Figure 4). Such ontology classification of comorbidities allows us to identify different groups of comorbidities and their hierarchical relations. Specifically, these comorbidities occur in various systems such as the cardiovascular system, blood, immune system, metabolism, digestive system, nervous system, kidney, respiratory system. Cirrhosis, viral hepatitis, and chronic hepatic failure belong to the digestive system. Dementia, seizure is the subclass of abnormal nervous system. Obesity and autoimmune deficiency status are also important risk factors for poor prognosis (Figure 4).

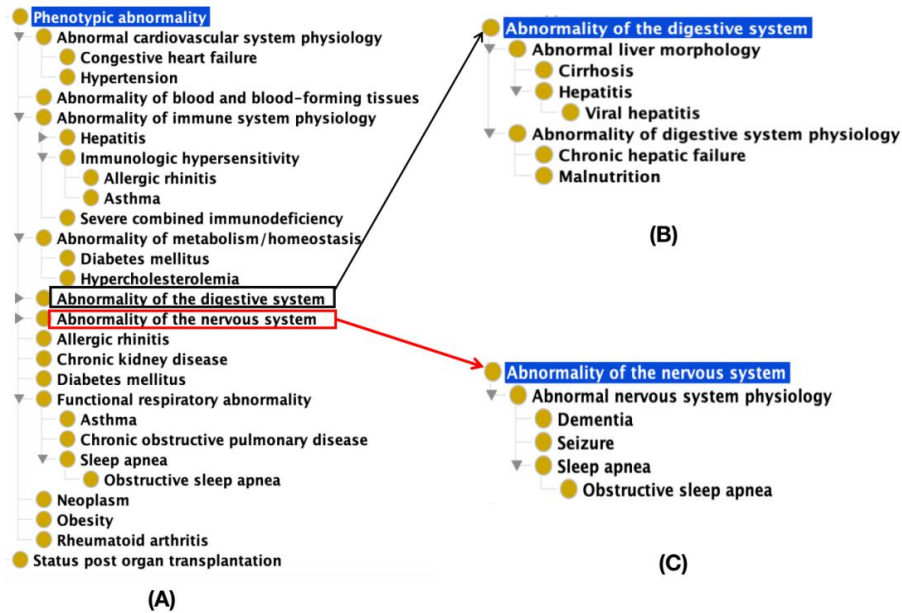

**Figure 4. Hierarchical lay of 23 common comorbidity phenotypes based on HP classification.**

*3.6. Correlation between the comorbidity phenotypes and severe disease outcomes*

To further study the relation between different comorbidity phenotypes and disease outcomes, we survey the disease data from the literature and compared the incidences of specific comorbidity phenotypes in severe or non-severe COVID-19 patients. From the long list of comorbidity phenotypes (Figure 4), we chose diabetes and kidney disease for our further analysis (Figure 5). The results from a total of 7 papers were applied for the study.

As shown in Figure 5, the morbidity of severe patients with diabetes or kidney failure phenotype was generally higher than that in non-severe patients. In all regions except California, the morbidity of kidney disease was higher in severe patients than in non-severe patients. In all records of COVID-19 patients, whether severe or non-severe, the incidence of diabetes was significantly higher than renal disease in COVID-19 patients. The incidence of diabetes was significantly higher in severe patients than that in non-severe patients (Figure 5). It was reported that cytokine storm might be activated in diabetic patients, leading to poor prognosis and death [3]. The exact mechanism deserves further investigation.

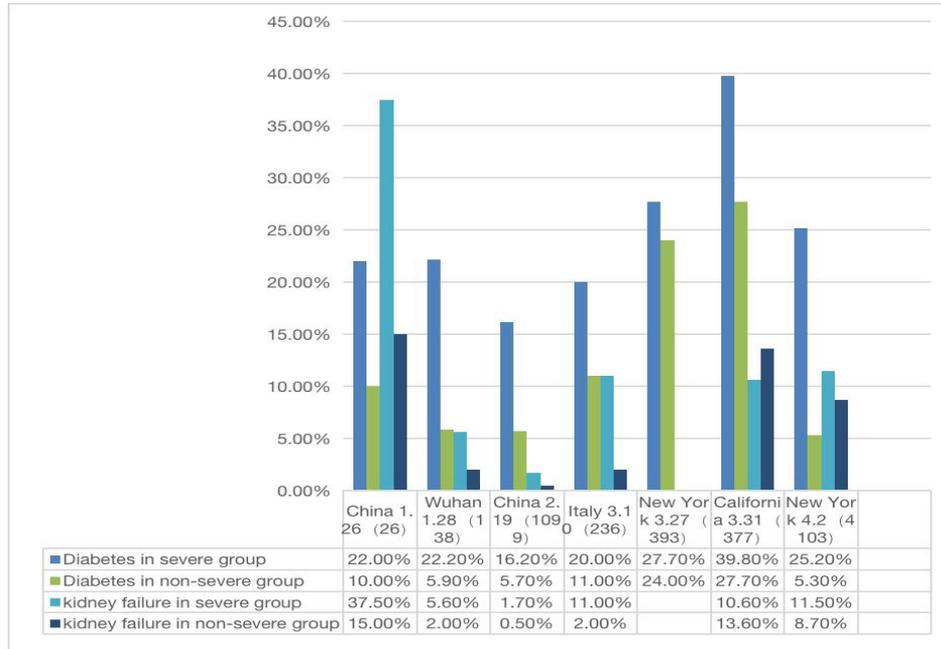

**Figure 5. Correlation between two comorbidity phenotypes (diabetes and kidney failure) and two disease outcomes (severe or non-severe).** In severe disease patients, the incidence of diabetes or kidney failure was higher than that in non-severe patient groups (The X-axis is country/city, report date, number of cases).

*3.7. CIDO ontological modeling of the COVID-19 phenotypes and comorbidities*

The Coronavirus Infectious Disease Ontology (CIDO; https://github.com/CIDO-ontology/cido) is a community-based ontology in the domain of coronavirus diseases with a specific focus on COVID-19. CIDO covers different coronavirus disease-related topics including coronaviruses, natural hosts of coronaviruses, phenotypes, comorbidities, drugs, and vaccines. To date, CIDO has more than 5,500 terms. In terms of COVID-19 phenotype classification, CIDO has imported the HP representations of COVID-19-related phenotypes (Figure 1) and comorbidities (Figure 4).

Different viruses have the disposition of inducing specific phenotypes in the patients. Currently study focuses on analyzing the COVID-19 phenotypes induced by SARS-CoV-2. CIDO also covers other coronaviruses such as SARS-CoV and MERS-CoV. To differentiate the relations between different viruses and phenotypes, we use a relation called '*pathogen susceptible to induction of phenotype*' defined in the Ontology of Host-Pathogen Interactions (OHPI) [14]. With this relation, we can define the virus-phenotype relation such as the following:

*SARS-CoV-2: 'pathogen susceptible to induction of phenotype' some headache*

## 4. Discussion

The worldwide epidemic of COVID-19 has caused serious damage and posed a serious threat to people's health and lives. Many specific drugs are still in clinical trials and vaccines are being developed in various countries. With the outbreak of COVID-19, many different clinical phenotypes are occurring in patients, so we want to map these phenotypes and comorbidities, to help doctors and CDC scientists better to know the clinical profiles of COVID-19.

Ontologies such as HPO and CIDO provides computer-interpretable bioinformatics resources for the analysis of phenotypes and underlying causes. HPO provides not only a standard phenotype terminology but also a collection of disease-phenotype annotations. CIDO reuses HPO and focuses on the identification of the causal relations between phenotypes and coronaviruses. These ontologies, together with ontology-based software programs and computational algorithms, can be combined to analyze the large amounts of data including clinical cases and basic experimental data with an aim to fully understand the internal mechanisms under different phenotypes of COVID-19 and their relations with various genetic mutations in the viruses, and to support rational development of therapeutic and preventive measures against the pathological infections.

Based on the HP classification, we systematically analyzed 17 clinical phenotypes of COVID-19 in case reported. We focused on three nervous phenotypes (loss of smell, loss of taste, and headache) and four abdominal phenotypes (nausea, vomiting, abdominal pain, and diarrhea) identified in patients, and found that patients from Europe and USA turned to have higher nervous phenotypes and abdominal phenotypes than patients from Asia. A total of 23 comorbidities were found to commonly exist among COVID-19 patients, usually COVID-19 patients with comorbidities had worse outcomes. From the study, patients with these comorbidities such as diabetes and kidney failure had worse outcomes compared with those without these comorbidities. Whether these results are related to the distribution of mutated virus strains in different regions and populations will be our next research direction. We will also investigate whether and how other conditions (such as temperature and season) are risk factors to the disease and infections.

Recently, children infected with COVID-19 presented kawasaki-like disease manifestations and systemic inflammatory responses syndrome in the United States and Europe, which led to critical illness and even deaths [10; 15]. And other patients infected with covid-19 have skin lesions as the main symptom, especially in toes, which is known as COVID-19 toes [8]. These new occurring phenotypes deserve our keen attention and require further careful monitoring and analysis.